\documentstyle[12pt]{article}
\begin{document}
\title{The spectrum and strong couplings of heavy-light hybrids}
\author {{\small Tao Huang$^{1,2}$\thanks{huangt@hptc5.ihep.ac.cn}, Hongying
Jin$^2$\thanks{jhy@hptc5.ihep.ac.cn} and Ailin
Zhang$^2$\thanks{zhangal@hptc5.ihep.ac.cn}}\\
{\small $^1$ CCAST (World Laboratory), P. O. Box 8730, Beijing, 100080}\\ 
{\small $^2$ Institute of High Energy Physics, P. O. Box 918, Beijing, 100039, 
P. R.China}\\}
\date {}
\maketitle
\begin{center}
\begin{abstract}
The spectrum of the $0^{++}$, $0^{--}$, $1^{-+}$ and $1^{+-}$ heavy-light
hybrids have been calculated in HQET. The interpolated current of the hybrid
is chosen as $g\bar q\gamma_{\alpha}G_{\alpha\mu}^aT^ah_{\it v}(x)$,
$g\bar q\gamma_{\alpha}\gamma_{5}G_{\alpha\mu}^aT^ah_{\it v}(x)$
and $g\bar q\sigma_{\mu\alpha}G_{\alpha\mu}^aT^ah_{\it v}(x)$.
Some strong couplings and decay widths of the heavy-light hybrids to
$B(D)\pi$ are calculated by using the QCD sum rules.
The mass of $0^{++}$ hybrid with gluon in TM($1^{--}$) or TE($1^{+-}$) mode
is found similar, while their decay widths are different. A two-point
correlation function between the pion and vacuum is employed  and the leading
order of $1/M_Q$ expansion is kept in our calculation.
\end{abstract}
\end{center}
\section{Introduction}
\indent
\par It is almost twenty years to search the exotic hadrons such as
the glueballs and hybrids. There are some special states which are regarded
as the candidates of hybrids, especially the $\hat \rho(1400)$ and
$\hat \rho(1600)$ have been studied widely, but no
confirmation has been made so far. Recently, these two special states arouse
great interest again. The E852 Collaboration at
BNL\cite{bnl} has reported a $J^{pc}=1^{-+}$ isovector resonance
$\hat{\rho}(1405)$ in the reaction $\pi^-p\rightarrow\eta\pi^0n$. They
also reported the mass $1370\pm 16^{+50}_{-30}$
MeV and width $385\pm 40^{+65}_{-105}$ MeV. The Crystal Barrel
Collaboration has also claimed to find an evidence
 in $p\bar p$ annihilation which may be resonance with a mass of
$1400\pm20\pm20$ MeV and a width of $310\pm50^{+50}_{-30}$ MeV\cite{bnl}.
The confirmation of these states will provide some evidence for the
existence of hybrids. At present, all the experiences specialize on the
light quark hybrids for some reasons, but it's necessary to extend the
energy region to hybrids including b or c quark, which is also possible in
the B or $\tau-C$ factory.   
\par Theoretically, the spectrum and decay width of hybrids have been
calculated widely with many methods such as bag model\cite{bag}, flux-tube model\cite{ft}, QCD sum rules\cite{sum}, lattice
\cite{latt} and other models\cite{other}. However, there are few works
about the spectrum and decay width of heavy-light hybrids\cite{grw}.
HQET has led to much progress in the theoretical understanding of the
properties of hadrons\cite{hqet}, one may ask a question whether it is
suitable in the heavy-light hybrids. For the heavy-light hybrids, the sum
rules' calculation in full QCD theory in
Ref.\cite{grw} shows that the component of gluon gives a contribution more than
$1.0 GeV$ to the mass, so the ``light freedom'' seems too heavy to keep the
$1/M_Q$ expansion available. The calculation of the spectrum of hybrids in
HQET will give an answer to the question. Our results show that the
spectrum of heavy-light hybrids including $b$ or $c$ quark are close to those
in full QCD theory, it is suitable to deal with these hybrids
within HQET. Besides, compared with $\bar bbg$ and $\bar ccg$ hybrid, the
spectrum and decay width of the heavy-light hybrid is easier to be
calculated in HQET.
\par It is interesting for the experimentalists to search the exotic $1^{-+}$
heavy-light hybrids, so theoretical determination of the property of these
states is necessary and urgent. In full QCD theory, the estimation in
Ref.\cite{grw} showed that the sum rule for the mass of $1^{-+}$ heavy-light
hybrids had no
platform at all. The masses of $1^{-+}$ hybrids were given under the main
assumption that the contribution of gluon condensate is less than $20\%$ of
the bare loop. In HQET, the ambiguous situation has been improved greatly. The
surface of the $\Lambda$ versus Borel variable $\tau$ varies little in a large
region, which gives a good platform and determines the masses of the hybrids.
\par According to the MIT bag model\cite{bcv}, the hybrids with the same
$J^{pc}$ have different internal interactions between the patrons which
indicate that they are different states, that is
to say, the gluon in hybrid can be in different TM($1^{--}$) or TE($1^{+-}$)
mode. In order to predict the properties of them, we should choose suitable
generating currents corresponding to these states for the calculation. In
the case of light 
quark hybrids\cite{sum}, these different $0^{++}$ states were found to have different
masses. In heavy-light hybrids case, the calculation shows that the mass split
of the $0^{++}$ heavy-light hybrids with gluon in different mode is not large
in the $1/M_Q \longrightarrow \infty$ limit. The mass of
$0^{++}$ hybrids from two different currents,
$g\bar q\gamma_{\alpha}G_{\alpha\mu}^aT^ah_{\it v}(x)$ with
gluon in TM($1^{--}$) mode and
$g\bar q\sigma_{\mu\alpha}G_{\alpha\mu}^aT^ah_{\it v}(x)$ with
gluon in TE($1^{+-}$) mode, is found similar.
\par Though the mass of these two different $0^{++}$ hybrids is similar,
their decay widths to $B(D)\pi$ final states are found different: 
the decay widths from current
$g\bar q\gamma_{\alpha}G_{\alpha\mu}^aT^ah_{\it v}(x)$ are about 86 MeV or
16 MeV corresponding to $B\pi$ or $D\pi$ final states, respectively, while
the ones from current
$g\bar q\sigma_{\mu\alpha}G_{\alpha\mu}^aT^ah_{\it v}(x)$ 
are 11 MeV or 2.6 MeV.
\par The decay widths of the $0^{++}$ and $1^{-+}$
hybrids to $B(D)\pi$ have been calculated in Ref.\cite{hjz1}, where the
decay constants of the hybrids were obtained from the formulae in full
theory\cite{grw}. Both the strong couplings and the decay constants are
calculated in HQET in this paper, and the strong coupling of the $0^{++}$
hybrid obtained here is much larger than that in Ref.\cite{hjz1}. The large
difference is from the values of the decay constant calculated in two ways.
\par In the calculation of the strong couplings, two-point function between
the pion and vacuum is taken use of
insteadly to avoid the ambiguity
resulting from the double Borel transformation for the three-point
correlation function and the infrared problem in soft pion limit. For
convenience, the calculation is kept in the leading order of $1/M_Q$ expansion.
\par The paper is organized as follows.  
The  analytic formalism of HQET sum rules for the spectrum of hybrid is given 
in Sec. 2. In Sec. 3, We give the numerical results of the spectrum and
decay constants of hybrid, the comparison of the spectrum with that in full
QCD theory is given too.
In Sec. 4, with the help of two-point correlation function between the pion
and vacuum, the analytic
formalism of HQET sum rules for the strong couplings of hybrids is derived
and the numerical results of some decay widths were obtained. In the last
section, we give the conclusion and discussion. 

\section{HQET sum rules for the spectrum of the heavy-light hybrid mesons}
\indent
\par As we know in Ref.\cite{sum}, the gluon in hybrid can be in different mode,
TM($1^{--}$) or TE($1^{+-}$) mode. To analyze these different hybrids, we
must choose suitable generating current. For the spectrum of the $0^{++}$
and $1^{-+}$ heavy-light hybrids with the gluon in TM($1^{--}$) and
TE($1^{+-}$) mode, respectively, the interpolated current in HQET is chosen as 
\begin{eqnarray}
j_{\mu}(x)=g\bar q\gamma_{\alpha}G^a_{\alpha\mu}T^ah_{\it v}(x).
\end{eqnarray}
where $q(x)$ is the light quark field and $h_{\it v}(x)$ is the heavy quark
effective field, $v$ is the velocity of the heavy quark. 
\par Then, we construct the correlation function as 
\begin{eqnarray}
\Pi_{\mu\nu}(\omega)
&=&i\int d^4x e^{iqx} \langle 0| T\{j_{\mu}(x),j^+_{\nu}(0)\}|0 \rangle \\\nonumber
&=&(v_\mu v_\nu -g_{\mu \nu})\Pi_v(\omega)+v_\mu v_\nu\Pi_s(\omega) ,
\end{eqnarray}
where
\begin{eqnarray}
\omega=2q\cdot v.
\end{eqnarray}

Since the free heavy quark propagator in HQET is 
${\it \int\limits^{\infty}_{0} d\tau\delta(x-v\tau)\frac {1+\rlap/v}{2}}$
and the interaction of the heavy quark with the gluon field $A_\mu$ in the
leading order of $1/M_Q$ expansion is $g\bar hv\cdot Ah$. Then under the 
fixed-point gauge $x_{\mu} A_{\mu}=0$(which will be used throughout this
paper), the full propagator of the heavy
quark $\langle 0|T(h(x)\bar h(0)|0 \rangle$ in the leading order of
$1/M_Q$ expansion is the same as the free one. The freedom of heavy quark
can be extracted out of the matrix element as a delta function, which
facilitates the calculation. 
\par In the operator product expansion, we keep the perturbative
term, two gluon condensate, three gluon condensate and two quark condensate.
The contribution of mixing condensate and higher dimension operators are
negligible since their smallness.The Feynman diagrams are shown in Fig. 1,
where the double line represents the propagator of the heavy quark. After
twice suitable Borel transformation, we obtain the $Im\Pi_s(\omega)$ and
$Im\Pi_v(\omega)$ corresponding to the scalar and vector contribution,
respectively 
\begin{eqnarray}
Im\Pi_s(\omega)& = &{\alpha_s \over 960\pi^2} \omega^6 
+ {\alpha_s \over 160\pi^2}m\omega^5 
- {1 \over 16}\langle \alpha_s G^2 \rangle \omega^2 
 - {m \over 8}\langle \alpha_s G^2 \rangle \omega \\\nonumber
& - &{\alpha_s \over 6} \langle {\bar qq}\rangle \omega^3
+ {\alpha_s \over 4}m \langle {\bar qq}\rangle \omega^2
- {\alpha_s \over 16} \langle gG^3 \rangle ,\\\nonumber
Im\Pi_v(\omega)& = &{\alpha_s \over 960\pi^2} \omega^6 
+ {\alpha_s \over 480\pi^2}m\omega^5 
+ {1 \over 48}\langle \alpha_s G^2 \rangle \omega^2 
 + {m \over 8}\langle \alpha_s G^2 \rangle \omega \\\nonumber
& - &{\alpha_s \over 18} \langle {\bar qq}\rangle \omega^3
+ {\alpha_s \over 4}m \langle {\bar qq}\rangle \omega^2
- {\alpha_s \over 48} \langle gG^3 \rangle ,
\end{eqnarray}
where the light quark mass corrections are considered also in these formulae. 
\par As for the $0^{--}$ and $1^{+-}$ hybrids, the current was chosen as
\begin{eqnarray}
j_{5\mu}(x)=g\bar q\gamma_{\alpha}\gamma_5G^a_{\alpha\mu}T^ah_{\it v}(x).
\end{eqnarray}
The correlation function expanded is similar to the vector current case
except for the
opposite sign for the contribution of the quark condensates and their
spectrum will be determined in a similar way.
\par For the $0^{++}$ hybrid with the gluon in TE($1^{+-}$) mode, the
following current should be used to predict the mass
\begin{eqnarray}
j(x)=g\bar q\sigma _{\mu\alpha}G^a_{\alpha\mu}T^ah_{\it v}(x).
\end{eqnarray}

The OPE of the $Im\Pi(\omega)$ for this current has been carried out as
\begin{eqnarray}
Im\Pi(\omega)= {\alpha_s \over 120\pi^2}\omega^6 
- m\langle \alpha G^2\rangle \omega
+ 2\alpha_s m \langle {\bar qq} \rangle \omega^2
- {\alpha \over 16}\langle gG^3\rangle .  
\end{eqnarray}
In the chiral limit, the two gluon and two quark condensates vanish.

On the phenomenal side, the decay constants of hybrids, $F_{H^\pm}$, are
defined as 
\begin{eqnarray}
\langle 0|j_{\mu}|H(0^{+})\rangle=F_{H^+}m^{1/2}_{H^+}v_\mu  ,
& \langle 0|j_{\mu}|H(1^{-})\rangle=F_{H^-}m^{1/2}_{H^-}\epsilon_\mu ,
& \langle 0|j|H'(0^{+})\rangle =F'_{H^+}m'^{1/2}_{H^+} .
\end{eqnarray}
where $m_H$ represent the masses of hybrids, $\epsilon_\mu$ is the
polarization vector and the two different $0^{++}$ hybrids with
gluon in TM($1^{--}$) or TE($1^{+-}$) mode is represented as
$H(0^{++})$ and $H'(0^{++})$, respectively. So the correlation function read
\begin{eqnarray}
\Pi_s(\omega) = -{F^2_{H^+} \over (2\Lambda-\omega)}
+\int_{\omega_c}^{\infty} d\omega'
{Im\Pi_s(\omega') \over \omega'-\omega} ,\\\nonumber
\Pi_v(\omega) = -{F^2_{H^+} \over (2\Lambda-\omega)}
+\int_{\omega_c}^{\infty} d\omega'
{Im\Pi_v(\omega') \over \omega'-\omega} ,\\\nonumber
\Pi(\omega) = -{F'^2_{H^+} \over (2\Lambda-\omega)}
+\int_{\omega_c}^{\infty} d\omega'
{Im\Pi(\omega') \over \omega'-\omega}.
\end{eqnarray}
where the first term of the right side is the pole term resulting from lowest
lying resonance contribution and the second term represents the contribution
of the continuum state and higher resonances, $\omega_c$ is the continuum threshold.
\par Taking use of the dispersion relations for the correlation function to
equate the both sides, we obtain
\begin{eqnarray}\label{spect}
{F^2_{H^+}\over (2\Lambda-\omega)} 
= &-{1\over\pi}\int_{0}^{\omega_c} d\omega' 
{Im\Pi_s(\omega') \over \omega'-\omega} ,\\\nonumber
{F^2_{H^-} \over (2\Lambda - \omega)} 
= &-{1 \over \pi}\int_{0}^{\omega_c} d\omega' 
{Im\Pi_v(\omega') \over \omega'-\omega} ,\\\nonumber
{F'^2_{H^+} \over (2\Lambda-\omega)} 
= &-{1\over\pi}\int_{0}^{\omega_c} d\omega' 
{Im\Pi(\omega') \over \omega'-\omega}.
\end{eqnarray}

After the Borel transformation\cite{neubert}, they are turned into 
\begin{eqnarray}
F^2_{H^\pm} e^{-2\Lambda/T}
= &-{1\over\pi}\int_{0}^{\omega_c} d\omega'
Im\Pi(\omega') e^{-\omega'/T} ,
\end{eqnarray}
where $T$ is the Borel transformation variable. 
So the $\Lambda$ can be determined as
\begin{eqnarray}\label{lam}
2\Lambda = {\int_{0}^{\omega_c}d\omega' \omega'Im\Pi(\omega')e^{-\omega'/T}\over \int_{0}^{\omega_c}d\omega' Im\Pi(\omega')e^{-\omega'/T}}
\end{eqnarray}

After the $\Lambda$ has been calculated, the decay constant can be
carried out according to (\ref{spect}).

\section{Numerical results of the spectrum and decay constants of the hybrids}
\indent
\par In this content, we will give the numerical results of the spectrum and
decay constants of the hybrids. To proceed the process, the mass of the b
and c quark are chosen as 4.7 GeV and 1.3 GeV, respectively, the condensates
are chosen as 
\begin{eqnarray}
\langle 0|m\bar qq|0 \rangle = -(0.1 GeV)^4 ,& \langle 0|\bar qq|0 \rangle =
-(0.24 GeV)^3,\\\nonumber
\langle 0|\alpha_s G^2|0 \rangle = 0.06 GeV^4 ,
&\langle 0|g G^3|0 \rangle = (0.27 GeV^2)\langle \alpha_s G^2 \rangle. 
\end{eqnarray}
and the scale of running coupling is set at the Borel parameter $T$.
\par The continuum threshold is chosen as below in the calculation:
$\omega_c=5.0$ GeV for the $0^{--}$ and two $0^{++}$ hybrids from current
$j_\mu(x)$ and $j(x)$, and
$\omega_c=4.5$ GeV for the $1^{+-}$ and $1^{-+}$ hybrids.

We display our results in HQET and those in full QCD theory in table. 1. In
this table, the right two columns represent the mass of heavy-light
hybrids calculated in full theory, the left represent the estimation
in HQET and the bottom of this table represents the $0^{++}$ hybrids
with the gluon in TE($1^{+-}$) mode.

\begin{table}
\caption{\label{tab1}
{\it  Masses of heavy-light hybrids with different $J^{pc}$}  (GeV).}
\begin{center}
\begin{tabular}{|c||c|c|c|c||c|c|}
\hline
&&&&&&\\
$J^{pc}$& $2\Lambda_c$ & $2\Lambda_b$ & $m_H(\bar qcg)$ & $m_H(\bar
qbg)$ & $m_c (full) $ & $m_b (full) $ \\
\hline
&&&&&&\\
$0^{++}$ & 4.4 & 4.4  & 3.5  & 6.9  & 4.0 & 6.8\\
\hline
&&&&&&\\
$0^{--}$ & 6.8  & 6.8  & 4.7  & 8.1 & 4.5 & 7.7 \\
\hline
&&&&&&\\
$1^{-+}$ & 3.6  & 3.6  & 3.1  & 6.5  & 3.2 & 6.3\\
\hline
&&&&&&\\
$1^{+-}$ & 3.8  & 3.8  & 3.2  & 6.6  & 3.4 & 6.5\\
\hline
&&&&&&\\
$0^{++}$ & 4.2  & 4.2  & 3.4  & 6.8 & none & none \\
\hline
\end{tabular}
\\
\end{center}
\end{table}

From the table, the mass of hybrids including b or c quark in HQET is
found similar to that in full theory, the light freedom in hybrids is not
large enough to break down the $1/M_Q$ expansion. So the calculation in HQET is suitable, which
implies that the $1/M_Q$ correction to the sum rules is not large.
\par In $1^{-+}$ hybrid case, the sum rule in full theory does not
stabilize\cite{grw}, which may indicate that no resonance exists in the
channel. The masses of these states were given under the main assumption that the
gluon condensate contribution is less than $20\%$ of the bare loop. In
HQET, the $2\Lambda$ of $1^{-+}$ hybrids versus Borel variable $\tau$ vary
little in a large region, which is shown in Fig. 2. The doted line
represents that of b quark hybrid and the real line represents that of c
quark, the little difference between them comes from the running coupling. The improvement of the
ambiguity may come from the reason\cite{neubert} that the T, $2\Lambda$ and
$\omega_c$ become constant low-energy
parameters in the $M_Q\to \infty$ limit in HQET, while the dependence of the
parameters $M^2$ and $s_c$ on the heavy quark mass is {\it priori} not
determined in full theory. Besides, the assumption in Ref.\cite{grw} is
proved reasonable here.
\par In the case of light quark hybrids case, the mass of $0^{++}$ hybrids
with the gluon in different mode was found to have a large
difference\cite{sum}. However, the $\Lambda$ for
the $0^{++}$ heavy-light hybrids with the gluon in different mode is found
similar in the $M_Q\to\infty$ limit. The mass of the heavy-light hybrids
in HQET is represented approximately 
\begin{eqnarray}
m\approx M_Q+\Lambda+O(1/M_Q),
\end{eqnarray}
so the mass split of the $0^{++}$ heavy-light hybrids is not large in HQET.
The mass of $0^{++}$ hybrid with gluon in TM($1^{--}$) mode is about 6.9 GeV
and 3.5 GeV corresponding to b or c quark hybrid, respectively, and the
mass of $0^{++}$ hybrid with gluon in TE($1^{+-}$) mode is about 6.8 GeV and
3.4 GeV, respectively.
\par When the radiative effects are taken account of, the effective current
would receive renormalization improvement and the heavy quark expansion of
the full current is necessary. However, in our derivation, neither the
radiative effects nor the $1/M_Q$ correction is taken account into.
\par The decay constants of the hybrids defined above can be obtained
through formula.(\ref{spect}), they are all collected in table. 2. The table
shows that the decay constants of the $0^{++}$ hybrid with gluon in
TE($1^{+-}$) mode are larger than those with gluon in TM($1^{--}$) mode. 
\begin{table}
\caption{\label{tab2}
{\it  Decay constants of heavy-light hybrids} ( $GeV^{7/2}$ ).}
\begin{center}
\begin{tabular}{|c||c|c|c|}
\hline
&&&\\
hybrid & $F_{H^+}(TM)$ & $F_{H^-}$ & $F'_{H^+}(TE)$  \\
\hline
&&&\\
$c$ quark & 0.31  & 0.28  & 0.97   \\
\hline
&&&\\
$b$ quark & 0.33  & 0.29  & 1.01  \\
\hline
\end{tabular}
\\
\end{center}
\end{table}

\section{Strong couplings and decay widths of heavy-light hybrids}
\indent
\par In Ref.\cite{hjz1}, we have calculated the decay width of
\begin{eqnarray}
\label{processa}
H_b(0^{++})(k) \longrightarrow B(0^{-+})(k-q) + {\it \pi}^{\pm}(q),\\
\label{processb}
H_b(1^{-+})(k) \longrightarrow B(0^{-+})(k-q) + {\it \pi}^{\pm}(q),
\end{eqnarray}
where the $0^{++}$ and $1^{-+}$ hybrids with gluon in the TM($1^{--}$) mode
and TE($1^{+-}$) mode, respectively, and the two-point correlation function
results from current $g\bar q\gamma_{\alpha}G_{\alpha\mu}^aT^ah_{\it v}(x)$.
The electric charges of the mesons except for pion have not been written out
explicitly. The cases of $H_c(0^{++}) \rightarrow D {\it \pi}^{\pm}$ and
$H_c(1^{-+}) \rightarrow D {\it \pi}^{\pm}$ have also been calculated there.
\par In this section, we will re-consider the same process in HQET firstly. 
For the decay widths of these processes, it is usually calculated through
the three-point vertex function or QCD light-cone sum
rules. However, in order to avoid the ambiguity of the three-point function
resulted from the double Borel transformation and the infrared divergence in
the soft pion approximation, we use the following two-point correlator
between pion and vacuum 
\begin{equation}\label{correlation}
A_{\nu}(\omega',\omega,v)=i\int dx e^{ikx}\langle {\it
\pi^{\pm}(q)}|T\{j_{1\nu}(x),j_2(0)\}|0
\rangle=A(\omega',\omega)v_{\nu}+B(\omega',\omega)(-q_{\nu}+q\cdot
vv_{\nu})
\end{equation}
where $j_{1\nu}(x)=g\bar q\gamma_{\mu}G^a_{\mu\nu}T^ah_{\it v}(x)$,
$j_2(x)=\bar h_{\it v}\gamma_5 q(x)$. $A(\omega',\omega)$ and
$B(\omega',\omega)$ are scalar functions of 
$\omega=2k\cdot v$ and $\omega'=2(k-q)\cdot v$
, which are determined through the spectral density saturated by the mesons
corresponding to the interpolated currents, respectively. The detailed OPE
expansion of the $A(\omega',\omega)$ and
$B(\omega',\omega)$ has been carried out in Ref.\cite{hjz1}.

\par For the the $0^{++}$ hybrid with gluon in TE($1^{+-}$) mode, the current
$j_{1\nu}(x)$ in the correlation function above should be replaced by 
$j'_1(x)=g\bar q\sigma_{\mu\alpha}G^a_{\alpha\mu}T^ah_{\it v}(x)$.
Then for the processes (\ref{processa}), we have another correlation function  
\begin{equation}
C(\omega',\omega)=i\int d^4x e^{ikx}\langle \pi^{\pm}(q)|T\{j'_1(x),j_2(0)\}|0
\rangle.
\end{equation}

\par In the infinite heavy quark mass limit, the following approximate
relation
\begin{equation}
2\Lambda-2\Lambda'\approx\omega-\omega'=2q\cdot v,
\end{equation}
will be used in this paper. where  $\Lambda \sim m_H-M_Q$ and $\Lambda' \sim
m_{meson}-M_Q$. Taking into account both the single pole terms and the
double pole term in the physical side, we can express 
$A(\omega',\omega)$, $B(\omega',\omega)$ and $C(\omega',\omega)$
functions of the single variable $\omega'$, respectively
\begin{eqnarray}
\label{a}
A(\omega')=\frac{F_{H^+}f_mg_{H^+m\pi}m^{1/2}_{H^+}m^2_m}{(2\Lambda'-\omega')^2M^3_Q}+{c_0
\over {2\Lambda'-\omega'}},\\
B(\omega')=\frac
{F_{H^-}f_mg_{H^-m\pi}m^{1/2}_{H^-}m^2_m}{(2\Lambda'-\omega')^2M^3_Q}+{c_1 \over
{2\Lambda'-\omega'}},\\
\label{b}
C(\omega')=\frac{F'_{H^+}f_mg'_{H^+m\pi}m^{1/2}_{H^+}m^2_m}{(2\Lambda'-\omega')^2M^3_Q}+{c_2
\over {2\Lambda'-\omega'}},
\label{c}
\end{eqnarray}
where $c_0$, $c_1$ and $c_2$ are constants. $F_i$ are decay constants of
hybrids defined above. $f_m$ are decay constants of the B or D meson and
$g_{H^\pm m\pi}$ refers to strong couplings, they are defined as below
\begin{eqnarray}
\langle 0|j_D| D\rangle = -if_Dm^2_D/M_c & ,
& \langle 0|j_B| B\rangle = -if_Bm^2_B/M_b ,\\\nonumber
\langle \pi^{\pm} (q) D|H'(0^{++})\rangle =g'_{H^+D\pi} & ,
&\langle \pi^{\pm} (q) B|H'(0^{++})\rangle =g'_{H^+B\pi} ,\\\nonumber
\langle \pi^{\pm} (q) D|H(0^{++})\rangle =g_{H^+D\pi} & ,
&\langle \pi^{\pm} (q) D|H(1^{-+})\rangle =g_{H^-D\pi}\epsilon\cdot q
,\\\nonumber
\langle \pi^{\pm} (q) B|H(0^{++})\rangle =g_{H^+B\pi} & ,
&\langle \pi^{\pm} (q) B|H(1^{-+})\rangle =g_{H^-B\pi}\epsilon\cdot q .
\end{eqnarray}

The formulae (\ref{a}) and (\ref{b}) is a little different from those in
Ref.\cite{hjz1} because of the different definition of the decay constants
of the hybrids.
\par Taking use of the dispersion relation and making Borel transformation
on them, we will get some equation about the strong couplings. After 
eliminating the $c_i$ terms with appropriate differentiation, these strong
couplings are obtained  
\begin{eqnarray}\label{am}
g_{H^+m\pi} ={M^3_Q\over
F_{H^+}f_mm^{1/2}_{H^+}m^2_m}[2\Lambda'A'(\tau)+A_0]e^{2\Lambda'/\tau},\\\nonumber
g_{H^-m\pi} ={M^3_Q\over
F_{H^-}f_mm^{1/2}_{H^-}m^2_m}[2\Lambda'
B'(\tau)+B_0]e^{2\Lambda'/\tau}\label{am2},\\\nonumber
g'_{H^+m\pi} =-{M^3_Q\over
F'_{H^+}f_mm^{1/2}_{H^+}m^2_m}e^{2\Lambda'/\tau}[2\Lambda'C(\tau)+C_0],
\end{eqnarray}
where $A_0$ and $B_0$ have been given in Ref.\cite{hjz1}, while $C(\tau)$
and $C_0$ have the form as
\begin{eqnarray}\label{ab}
C(\tau)&=&8\sqrt{2}\{3[(m_H-m_m)b_2-2b_1]-(m_H-m_m)F_1/\tau\},\\
C_0&=&-8\sqrt{2}(m_H-m_m)F_1,
\end{eqnarray}
where the parameters $b_i$, $F_i$  have been calculated in Ref.\cite{hjz1} too.
\par Before going on the numerical calculation of the strong couplings and
decay widths, it is necessary to fixing the parameters firstly.
The masses of the heavy quarks and heavy mesons have been given in
Ref.\cite{epj}, the decay constants of B and D mesons are chosen as
Ref.\cite{rr}. The masses and decay constants of the heavy-light hybrids
have been computed above. The numerical results of the strong couplings
 are shown as $Fig. 3 \to Fig. 5$, where the value of them is
determined around $\tau\sim 3.0 GeV$. They are all displayed in table. 3,
where the $g_{H^-m\pi}$ is dimensionless. 
\begin{table}
\caption{\label{tab3}
{\it Some parameters input and strong couplings of hybrids. ( GeV ).}}
\begin{center}
\begin{tabular}{|c||c|c|c|c|c|c|c|c|c|}
\hline
&&&&&&&&&\\
hybrid & $M_Q$ & $m_m$ & $f_m$ & $m_H(0^{++})$ & $m_H(1^{-+})$ &
$m'_H(0^{++})$ & $g_{H^+m\pi}$ & $g'_{H^+m\pi}$ & $g_{H^-m\pi}$\\
\hline
&&&&&&&&&\\
b  & 4.7  & 5.28  & 0.18  & 6.9  & 6.5  & 6.6 & 8.5 & 3.2 & 2.8\\
\hline
&&&&&&&&&\\
c  & 1.3  & 1.87  & 0.19  & 3.5  & 3.1  & 3.2 & 2.0 & 0.8 & 0.8\\
\hline
\end{tabular}
\\
\end{center}
\end{table}
 
\par  To the processes (\ref{processa}) and (\ref{processb}), the decay
widths are given by the following formulae
\begin{eqnarray}\label{de}
\Gamma(H(0^{++})\rightarrow m(0^{-+}) + {\it \pi}) =&
{g^2_{Hm\pi} \over 8\pi}{|q| \over m^2_H} =&
{g^2_{Hm\pi}\over 16\pi}{m^2_H-m^2_m \over m^3_H},\\\nonumber
\Gamma(H(1^{-+})\rightarrow m(0^{-+}) + {\it \pi}) =&
{g^2_{Hm\pi} \over 24\pi}{|q|^3 \over m^2_H} =&
{(m^2_H-m^2_m)^3g_{Hm\pi}^2\over 192 \pi m^5_H}.
\end{eqnarray}

Then in the $M_Q\to\infty$, the numerical results of the decay widths read
\begin{eqnarray}
\Gamma(H(0^{++})\rightarrow B(0^{-+}) + {\it \pi}) = 86 MeV ,
&\Gamma(H(0^{++})\rightarrow D(0^{-+}) + {\it \pi}) = 16 MeV ,\\
\Gamma(H(1^{-+})\rightarrow B(0^{-+}) + {\it \pi}) = 2.2 MeV ,
&\Gamma(H(1^{-+})\rightarrow D(0^{-+}) + {\it \pi}) = 1.0 MeV .\\
\Gamma(H'(0^{++})\rightarrow B(0^{-+}) + {\it \pi}) = 11 MeV ,
&\Gamma(H'(0^{++})\rightarrow D(0^{-+}) + {\it \pi}) = 2.6 MeV .
\end{eqnarray}

Though the decay of hybrids appears to follow the $S + P$ selection rule,
which means that the decay of hybrids to two S-wave mesons are suppressed
\cite{page}, the selection rule is not absolute.
In the flux tube and constituent glue models, it can
be broken by wave function and relativistic effects, and the bag model predict
that it is also possible that the excited quark loses its angular momentum
to orbital angular momentum \cite{godfrey}, the results obtained here
support this idea.
\par The decay widths of the processes to $B(D)\pi$ final states for $H(0^{++})$
with gluon in TM($1^{--}$) mode is much larger than those for $H(1^{-+})$
with gluon in TE($1^{+-}$)
mode, the reason is that the final states in the later channels are
in the P wave. Besides, the decay width of the $H(0^{++})\to B\pi$ obtained
here is much larger than that we got in Ref.\cite{hjz1}, the difference is
from the decay constant. The decay constant of the $0^{++}$ hybrid we got
there in full theory is much smaller than the one calculated above in HQET.
\par Though the mass of these two $0^{++}$ hybrids is almost the same,
the strong couplings to pion of them are different, so the decay
widths of these two different $0^{++}$ hybrids are different. The decay
width of the $H'(0^{++})\to B(D)\pi$ are smaller
than those corresponding to $H(0^{++})\to B(D)\pi$. The physical reason about
the difference of decay width of these two different $0^{++}$ hybrids is
unknown to us yet, but the difference between these states provides a nice
evidence that the decay property of these two $0^{++}$ hybrids with the
gluon in different mode is different. 
\section{Conclusion and Discussion}
\indent
\par We calculate the spectrum of the $0^{++}$, $0^{--}$, $1^{-+}$ and
$1^{+-}$ heavy-light hybrids with different currents in HQET, the results
from current $g\bar q\gamma_{\alpha}G_{\alpha\mu}^aT^ah_{\it v}(x)$
and $g\bar q\gamma_{\alpha}\gamma_{5}G_{\alpha\mu}^aT^ah_{\it v}(x)$ are
compatible to those in full QCD theory. The calculation shows that the light
freedom in heavy-light hybrids is not heavy enough to break down the $1/M_Q$
expansion and it is suitable to apply HQET to heavy-light hybrid systems. 
\par The sum rules for the masses of $1^{-+}$ heavy-light hybrids have no
platform at all in full theory, so the masses of them were given under some
assumptions. The ambiguity of these sum rules have been improved in HQET, which
suggests the reasonableness of the assumptions in Ref.\cite{grw} in another
way. In the calculation, the leading order $1/M_Q$ expansion approximation
is used and only the first two terms in OPE are kept in our estimate, which
will bring some errors. Besides, the decay constants will bring in
large deviation too, it is necessary to determine them more precisely. 
\par Since the gluon in hybrids can be in different mode, the hybrids with
the same $J^{pc}$ but different gluon mode are in fact different states. In
the case of light quark hybrids, the two different $0^{++}$ states have
different masses definitely. In the heavy-light hybrids case, the masses of
these two different states are found similar in the $M_Q\to\infty$ limit,
however, 
the calculation shows that the decay widths in the processes of these two
hybrids to $B(D)\pi$ final states are different. The decay width
of $H(0^{++})\to B(D)\pi$ is
found about 86(16) MeV, while the decay width of $H'(0^{++})\to B(D)\pi$ is
only 11(2.6) MeV.
\par The strong couplings and decay widths of the heavy-light hybrids in the
processes of $H(1^{-+})\to B(D)\pi$ are calculated too. The large difference
of the decay widths calculated here from those calculated in Ref.\cite{hjz1}
lies on the large difference between the decay constants calculated in two
different ways.\\

\vspace{1.0cm}
{\bf Acknowledgment}

This work is supported in part by the national natural science foundation 
of P. R. China.

\newpage
\par
{\huge\bf Figure caption}\\
\par
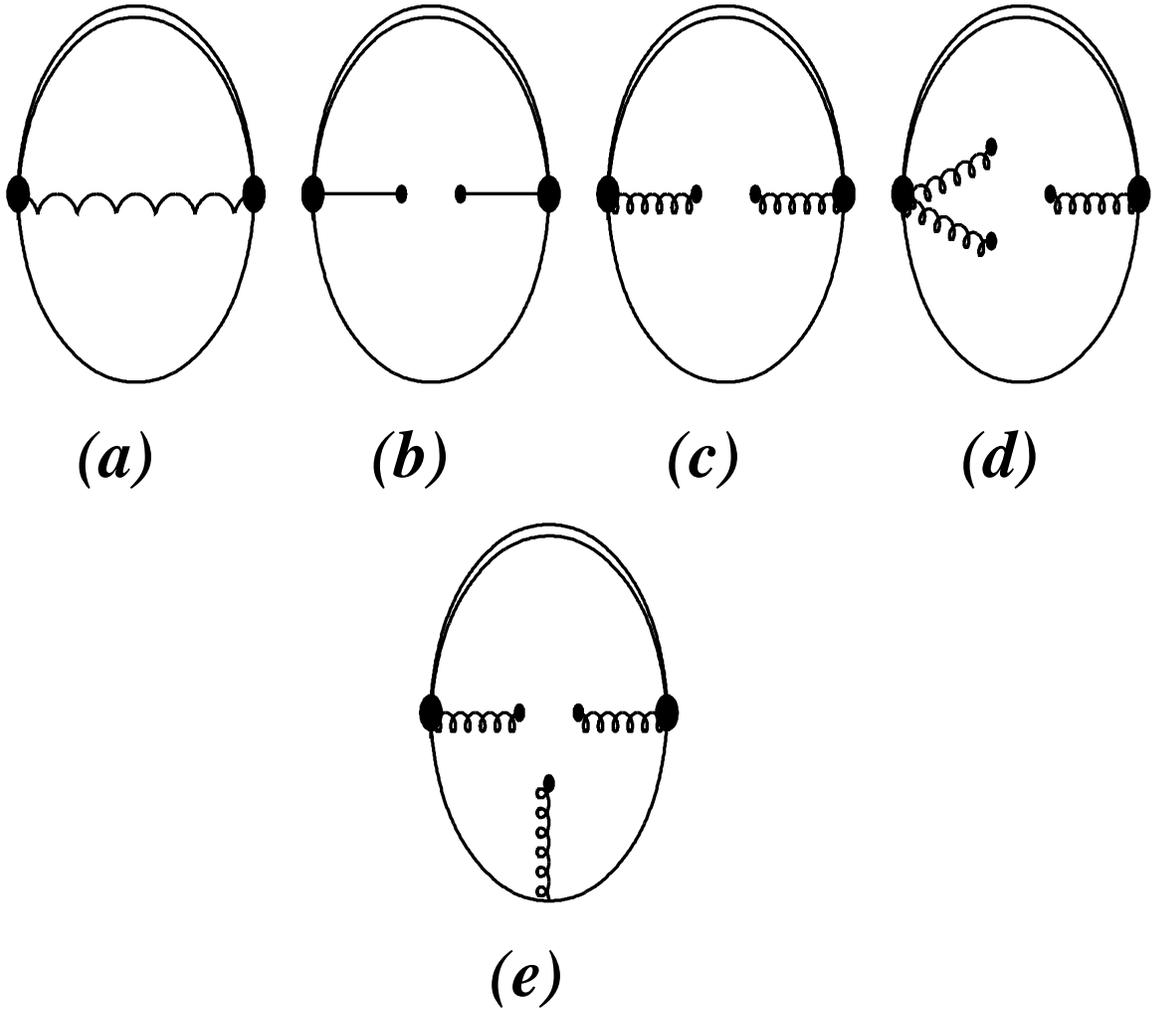
Fig. 1: Feynman diagrams contributing to the correlation function in HQET.\\
\par
\par
Fig. 2: $2\Lambda$ of the $1^{-+}$ heavy-light hybrids versus Borel variable T.\\
\par
\par
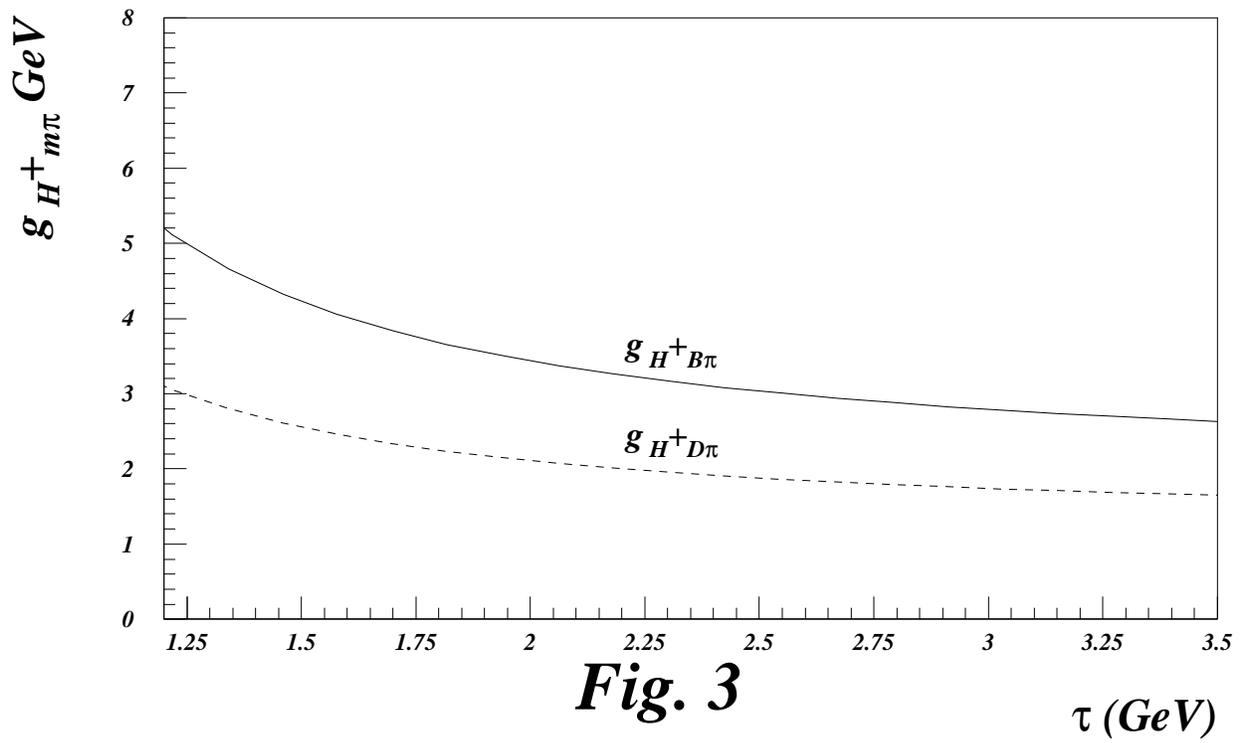
Fig. 3: Strong coupling of $H(0^{++})$ heavy-light hybrids versus Borel
variable $\tau$.\\
\par
\par
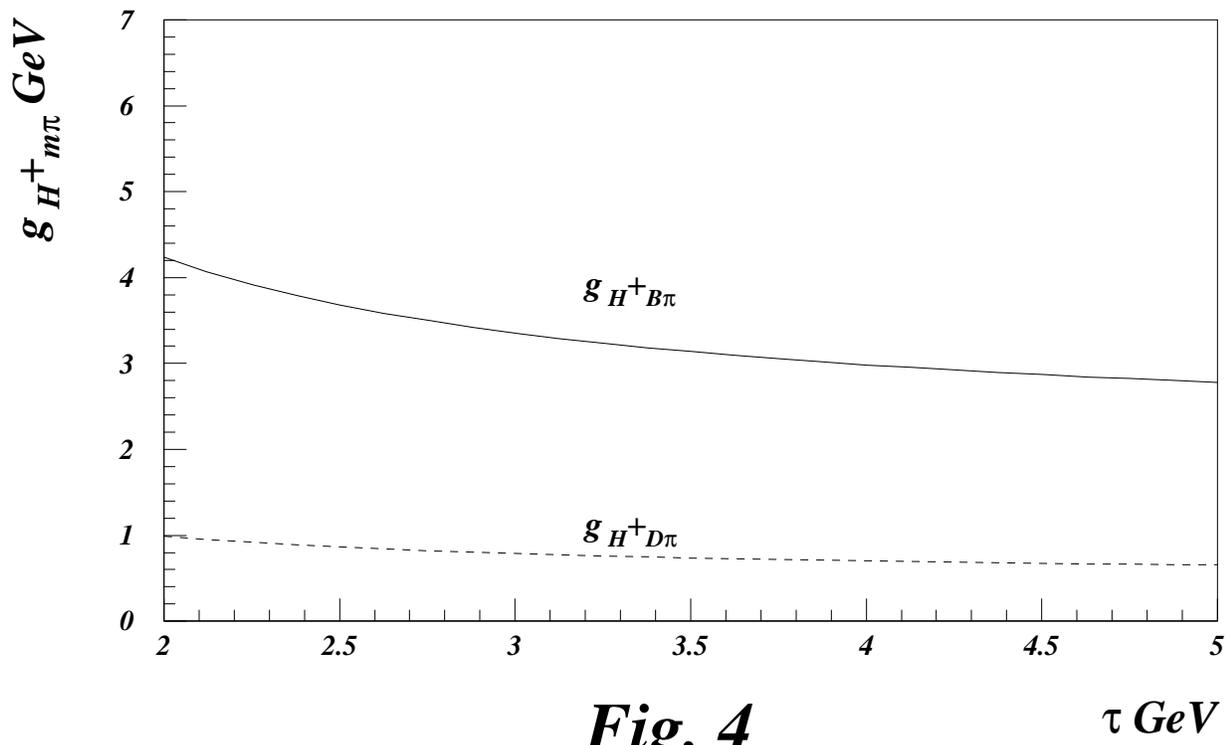
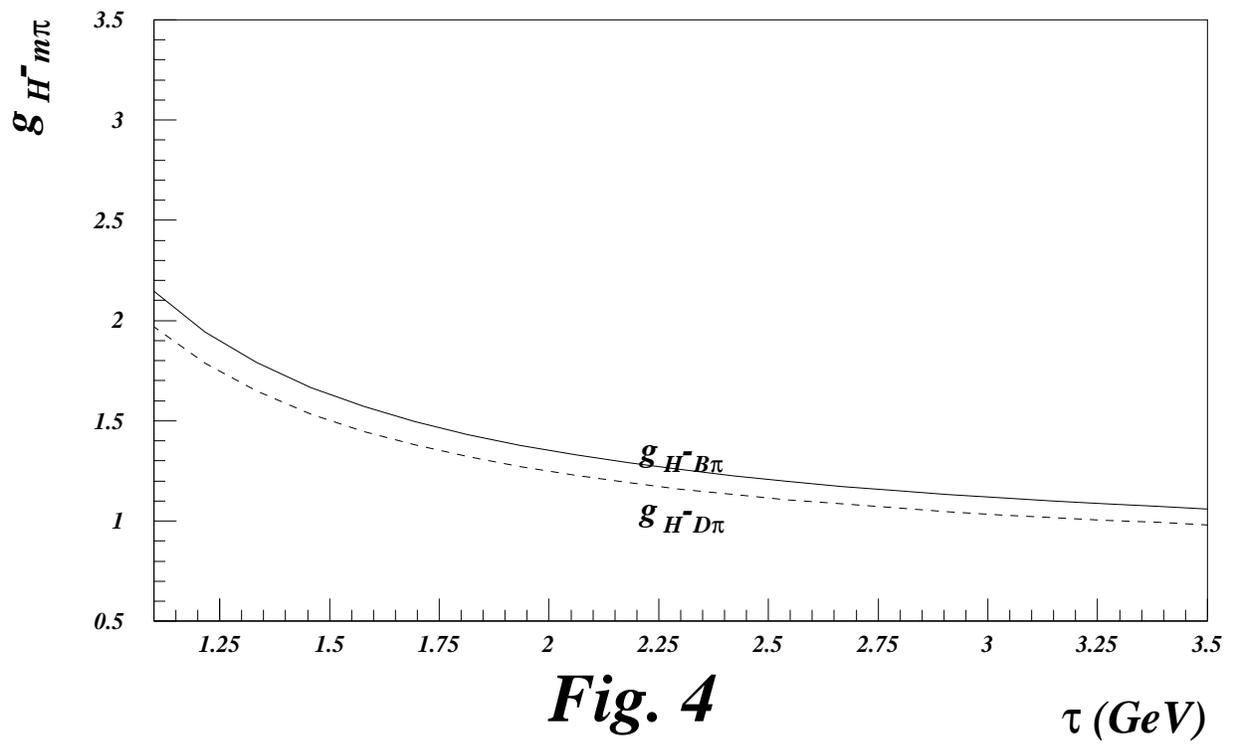
Fig. 4: Strong coupling of $H'(0^{++})$ heavy-light hybrids versus Borel
variable $\tau$.\\
\par
\par
Fig. 5: Strong coupling of $H(1^{-+})$ heavy-light hybrids versus Borel
variable $\tau$.\\
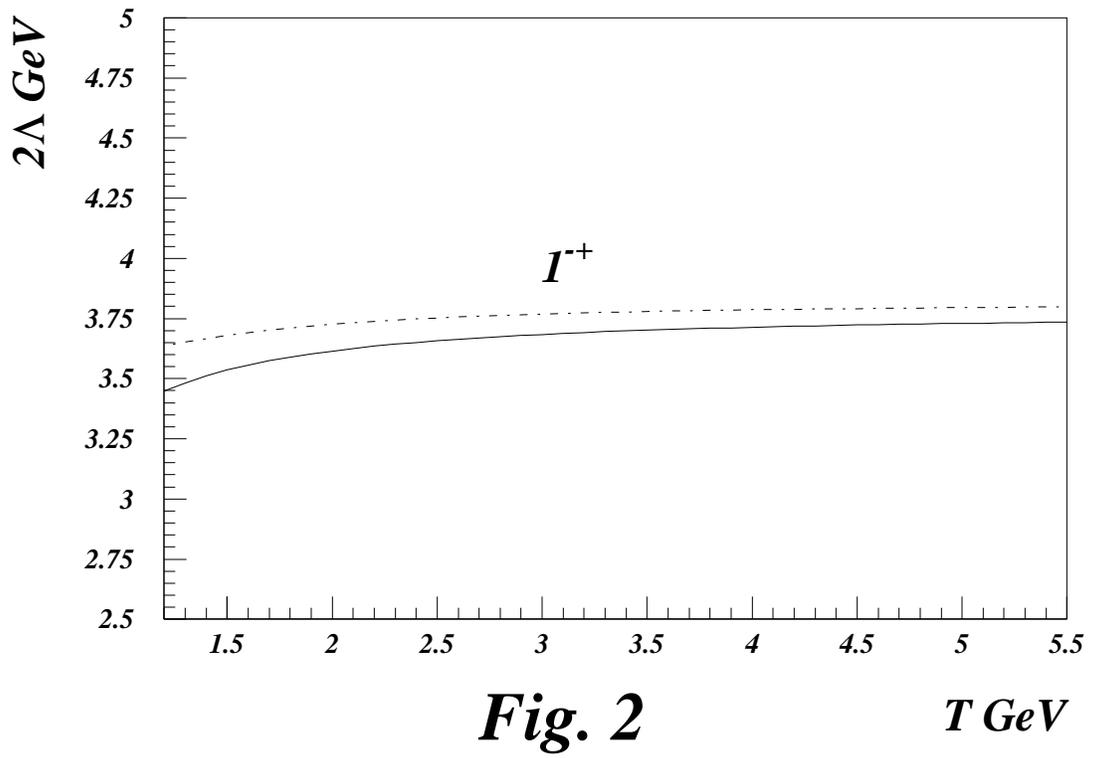
\end{document}